\documentclass[graybox, natbib, footinfo]{svmult}

\usepackage{mathptmx}       % selects Times Roman as basic font
\usepackage{helvet}         % selects Helvetica as sans-serif font
\usepackage{courier}        % selects Courier as typewriter font
\usepackage{type1cm}        % activate if the above 3 fonts are
\usepackage{makeidx}         % allows index generation
\usepackage{graphicx}        % standard LaTeX graphics tool
\usepackage{multicol}        % used for the two-column index
\usepackage[bottom]{footmisc}% places footnotes at page bottom

\makeindex             % used for the subject index
\begin{document}
\title*{4MOST - 4m Multi Object Spectroscopic Telescope}
\author{\'{E}ric Depagne and the 4MOST consortium}
\institute{\'{E}ric Depagne, Leibniz-Inisitut f\"{u}r Astrophysik Potsdam, Potsdam , Germany \linebreak \email{edepagne@aip.de}
%\and Name of Second Author \at Name, Address of Institute \email{name@email.address}}
}
\maketitle

\abstract{4MOST (4m Multi Object Spectroscopic Telescope) is a spectroscopic facility that will be installed on ESO's VISTA around 2020. The science rationale of this facility are to be found in the ASTRONET Science Vision for European Astronomy \citep{de2007}. Specifically fundamental contribution can be made to the Extreme Universe (Dark Energy \& Dark Matter, Black holes), Galaxy Formation \& Evolution, and the Origin of Stars science cases in the ASTRONET Science Vision. The unique capabilities of the 4MOST facility are due to by its large field-of-view, high multiplex, its broad optical spectral wavelength coverage}

\section{History}
\label{sec:history}
In 2010, ESO issued a Call for Letters of Interest (LoI) for the Conceptual Design of a Multi-Object Spectroscopic  instrument or facility. AIP, leading a consortium of 11 (at that time) institutes in Europe, answered this call by proposing 4MOST. The main science drivers for this proposal, was to allow a complete exploitation of the data to be provided to the astronomical community by two spacecrafts:
\begin{itemize}
\item[-] Gaia
\item[-] E-rosita.
\end{itemize}
The proposal was made with an unchosen location at that time, and the concept could be installed on both the NTT and VISTA. ESO then chose in may 2012 that the telescope of choice for such an instrument would be VISTA located at the Paranal Observatory. The Conceptual Design Phase was completed in May 2013, with the official selection by ESO, alongside the other project selected during this call for LoI : MOONS, which will be installed on the VLT. The offical starting date as an ESO project is january 2015, with an installation planned on VISTA in 2019.

\section{Science Drivers}
\label{sec:sciencedrivers}
The call for LoI made it explicit that the facility should allow ESO's astronomers to open up new areas in research that would be left unexplored otherwise. But let the proposer choose any other science goal they wish.

This lead the 4MOST consortium to define 6 Design Reference Surveys (DRS) that will be the main scientific drivers for the design of the facility, with the clear understanding that the decision they implied should not , as much as possible, hinder the possibility of other science to be done.
The 6 DRS are :
\begin{itemize}
\item[-] Milky Way Stellar Halo (High and Low Resolution)
\item[-] Milky Way Bulge and Disk (High and Low Resolution)
\item[-] Cluster of Galaxies (Low Resolution)
\item[-] AGN (Low Resolution)

\end{itemize}

The main constraint for these DRS, is that they have to be finished within 5 years. Finished, for the DRS has a special meaning. For each DRS, we have defined a Figure of Merit (FoM) that allows to measure the progress over time of the observations. For each DRS, we required that the FoM be 1 at the end of the 5 years. al

\subsection{Milky Way}
\label{subsec:milkyway}
There are of course many questions that 4MOST will allow to tackle, and it is not possible to list them all (since we know some will come up with ideas to use the data in ways we did not forsee), but here is a list of what we already plan to do:
\begin{itemize}
\item[-] Determine the MW 3D potential from streams to $\sim$ 100 kpc
\item[-] Measure the effects of baryons
\item[-] Map out the mass spectrum of Dark Matter halo, substructure by the kinematic effects
on cold streams of $10^3 - 10^5 \mathrm{M}_{\odot}$
\item[-]Identify moving groups in velocity distribution up to 10 kpc (Hipparcos did it only up to 200 pc). 4MOST will allow to go half-way through the Galaxy.
\item[-] Derive abundances on an unparalleled scale. The High Resolution channel will allow a determination down to a precision of 0.1 - 0.2  dex \citep{Caffau2013} for all the major nucleosynthesis channels:
\begin{itemize}
\item[$\bullet$] Light elements
\item[$\bullet$] $\alpha$ elements
\item[$\bullet$] Iron-peak elements
\item[$\bullet$] r-process elements
\item[$\bullet$] heavy and light s-process elements
\item[$\bullet$] odd Z elements (such as Na, Al)
\end{itemize}
\item[-] Observe $10^5$ halo stars, leading to the most accurate determination of the Metallicity Distribution Function, allowing the identification of chemo-dynamical substructure (such as streams of tidally disrupted dwarves galaxies) and most importantly, understanding for those substructure, if the formation is done in-situ of mostly by accretion.
\item[-] Observe $1.5\times10^5$ bulge giants, allowing to understanding the effects of reddening and the existence of substructure (with a very good overlap with what MOONS will be doing at the VLT), disantangle the bulge formation scenarios : Collapse or merging of proto-galaxies
\end{itemize}

%Use the standard \verb|equation| environment to typeset your equations, e.g.
%%
%\begin{equation}
%a \times b = c\;,
%\end{equation}
%%
%however, for multiline equations we recommend to use the \verb|eqnarray| environment\footnote{In physics texts please activate the class option \texttt{vecphys} to depict your vectors in \textbf{\itshape boldface-italic} type - as is customary for a wide range of physical subjects}.
%
\subsection{AGN and Glaxies Clusters}
\label{subsec:AGN}

\section {Data policy}
\label{sec:datapolicy}

The raw data gathered by 4MOST during its lifetime (which is currently expected to be around $10$ years) will be made immediately public. The calibrations will also be made public, so that anyone interested in reducing and using the data will have the opportunity to do so. But the 4MOST consortium will develop in-house a data reduction pipeline, that will produce not only calibrated spectra, but also an analysis of those spectra, to produce a catalog that will contain, for each object, the stellar parameters, but also a chemical analysis.

\section{Concept Design}
\label{sec-conceptdesign}

4MOST is an instrument that will be installed on VISTA. It will be available for observations 100\% of the time (which is a major difference with instrument of the same kind to be installed at other observatories).

The main characteristics of the instruments are as follows:
\begin{itemize}
\item[-] Large Field of View. The current design allows for a FoV of 2.5$^{\circ}$ in diameter, providing an area of 4.06 $\mathrm{deg}^{2}$ on sky. The design of the correcting optics (that includes an Atmospheric Dispersion Corrector) allows for the whole field of view to be seeing limited.
\item[-] Fibre Positioner. The current design of this critical part of the system, is for an Echidna-style positioner. An Echidna-style positioner (named AESOP) allows a quick set-up of the fibres on the focal plane, opening the possibility to reconfigure the spines on a short timescale (basically, it can be done every 20mn). It allows an efficient coverage of the FoV, and a high density of fibres. The fibre pitch (describing how far from the center of a spine, the next spine center is located) is as small as 10mm, with a patrol radius (indicating how far from its center a spine can move), is 12mm. The accuracy of the positioning of each spine is very good and allows for 99\% of the spines to be located within $0.1''$ of their requested position. The current design allows 2400 fibres to be positioned on the focal surface.
\item[-] Spectrographs. 4MOST will observe simultaneously in its two spectroscopic channels a High-Resolution and a Low-Resolution. The details of each channel can be found in table \ref{tab:highlowresSG}
\begin{table}
\caption{Details of each spectroscopic channel}
\label{tab:highlowresSG}
%
% Follow this input for your own table layout
%
\begin{tabular}{cc|c}
\hline\noalign{\smallskip}
&High Resolution & Low Resolution    \\
\noalign{\smallskip}\svhline\noalign{\smallskip}
Number of fibers    & 800   & 1600 \\
Resolution          & $> 20000$& $>5000$ \\
Wavelength coverage& 390 - 457 & 390-950 nm \\
& 595 - 950 & \\
\noalign{\smallskip}\hline\noalign{\smallskip}
\end{tabular}
\end{table}

\end{itemize}

\subsection{Expected Performaces}
\label{subec:expecteedperfs}
\begin{figure}[h]
\sidecaption
% For example, with the graphicx style use
\includegraphics[scale=.3]{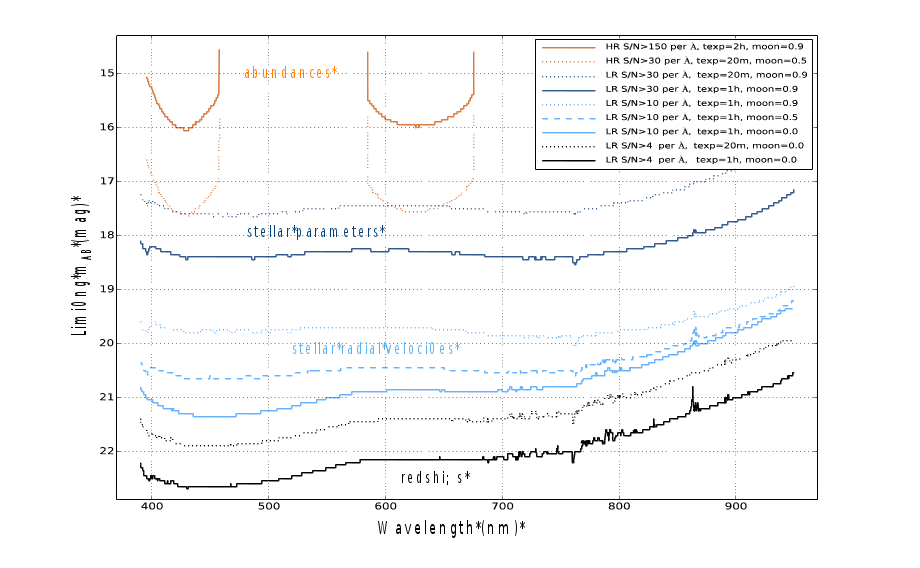}
%
% If no graphics program available, insert a blank space i.e. use
%\picplace{5cm}{2cm} % Give the correct figure height and width in cm
%
\caption{This figure shows the magnitude as a function of wavelength, that can be reached for various scenarios, by 4MOST. This plot shows how efficient this facility will be, and how versatile it will be, allowing to tackles key astrophysical problems both in Galactic and Extra-galactic Astrophysics.}
\label{fig:signaltonoise}       % Give a unique label
\end{figure}

\section {4MOST Facility Simulator (4FS)}
\label{sec:4mostperfsimul}

\subsection{Description}
\label{subsec:4mostdescr}
The 4MOST Facility Simulator is a set of tools that we use in the project to ensure the science goals will be reachable. It has been designed to be very versatile, allowing us to check the impact of design changes, but also observing strategy changes. Thus, it has become a crucial tool for the evolution of the project. For an ease of use, and to ensure a very good efficiency of the simulator, it has been designed as a modular tool, comprised of the following 3 main parts:
\begin{itemize}
\item[-] a throughput simulator (TPS)
\item[-] a data quality control tool (DQCT)
\item[-] and an operation simulator tool (OpSim).
\end{itemize}
The input for the 4FS consists in the catalogs that have been prepared for each DRS. These catalogs contain a list of objects, and among other informations, their coordinates, their magnitudes their type and the requested oberving time to consider the observation a success. Along these catalogs comes a series of spectral templates that will be used to produce a simulated observation with 4MOST.

\subsection{Workflow}
\label{subsec:4mostworkflow}
The 4FS Workflow has been devised in order to maximise the efficiency of the simulations. We describe it briefly in the coming section.

\begin{itemize}
\item[-] TPS :
\begin{itemize}
\item[$\bullet$] The first step is to fold a noise-free template spectrum with a simulated instrument response, in order to obtain a noisy "observed" spectrum. This is done for a variety of exposure times : from 10mn to 240mn.
\item[$\bullet$] Then a simulated sky background, corrsponding to the three observing period we have defined : bright-, grey- and dark-time is added.
\item[$\bullet$] This procedure is repeated in 1 magintudes step, to cover the whole magnitude range of the expected observed targets. This leads to approximatively $4000$ templates to analyse.
\end{itemize}
\item[-] DQCT :
\begin{itemize}
\item[$\bullet$] Once the simulated spectra have been produced, the Signal to Noise ratio for each exposure is computed
\item[$\bullet$] According to this S/N evaluation, the exposure time for each template is computed, and this parameter is included in the Operations Simulator.
\end{itemize}
\item [-] OpSim
\begin{itemize}
\item[$\bullet$] With the updated list of targets, a new list of targets to be observed is produced, and a new field configuration is generated. The process is repeated until the first 5 years of operations are over.
\end{itemize}
\end{itemize}


\subsection{Survey Strategy}
\label{subsec:surveystrategy}

We have defined various possibilities for the 4MOST facility, with different number of fibres, spectrographs, etc.
We have defined what we call a "Baseline" facility, which is defined as follows (the Moon phases and the predominant Northern Winds are taken into account):
\begin{itemize}
\item[-] AESOP Fibre positioner, 1624 LR + 812 HR fibres, 4.06 deg2 FoV (hex), 5\% overlap between tiles, 100 (LR) + 50 (HR) sky fibres reserved
\item [-] Exposure times: everywhere 6 x 20 min, more repeats on Bulge and fewer (3 x 20 min) in selected areas on-Disk
\item [-] 1827 nights, with 5 (sched.) + 6 (unsched.) nights of technical downtime p.a., and 20% nights lost to weather

\end{itemize}
\begin{figure}[h]
\sidecaption
% For example, with the graphicx style use
\includegraphics[scale=.3]{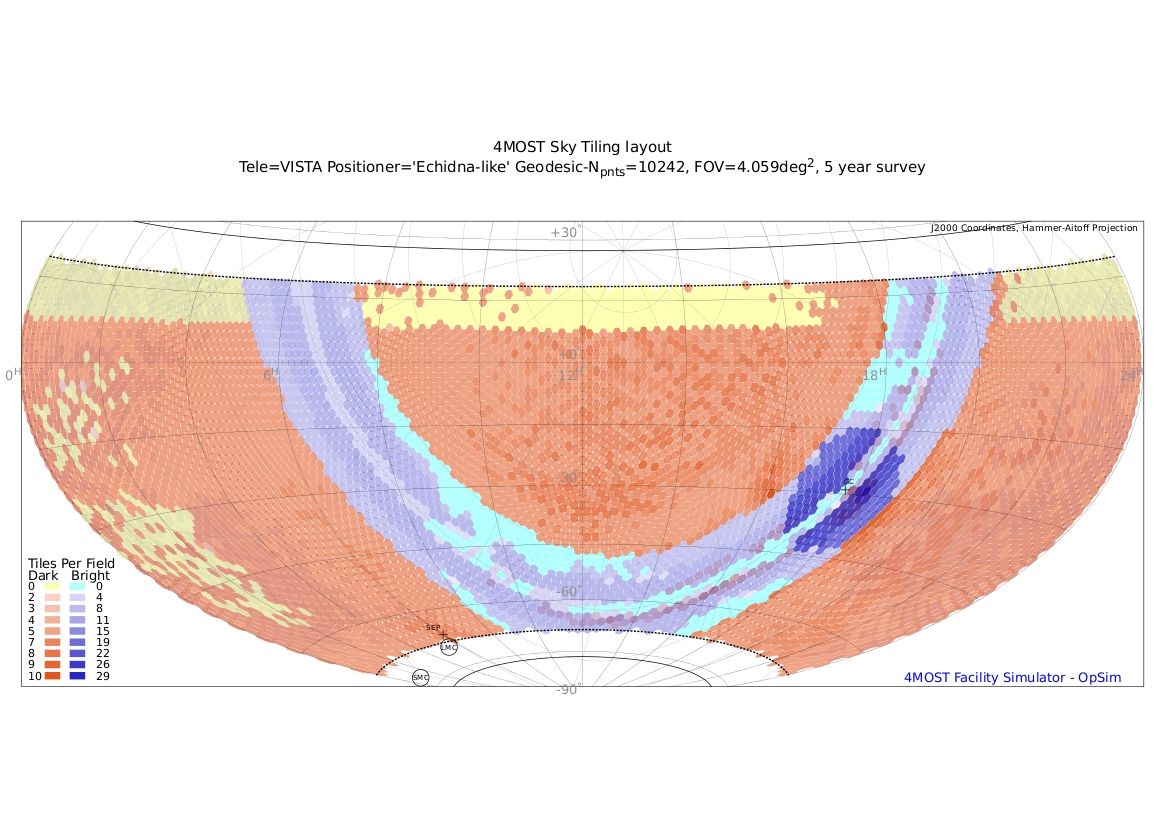}
%
% If no graphics program available, insert a blank space i.e. use
%\picplace{5cm}{2cm} % Give the correct figure height and width in cm
%
\caption{This figure shows the observed sky during the first 5 years of the survey. The color code indicates how frequently each tile has been observed. The whole sky is covered at least once, but there are important parts of the sky, such as the South Galactic Pole, which are seriously underobserved. This will be fixed with a more sophisticated Observing Strategy, but the RA pressure on this region is significant.}
\label{fig:tilingexecuted}
\end{figure}

\section {Observed Targets and Fiber efficiency}
The number of targets that will be observed by 4MOST during the first 5 years of operations is round 35 millions, all DRS included. The computation of the observed target is made using the Baseline, defined previously, setting for 4MOST. Table \ref{tab:obstarg} gives the number of targets for each individual DRS while table \ref{tab:obsefficiency} summarizes the predicted efficiency of 4MOST.
\label{sec:observedtargets}
\begin{table}
\begin{center}
\begin{tabular}[c]{lrrr}
DRS & Observed targets & Successful targets & Success rate\\
\hline
Halo LR & 1746500 & 150100 & 0.86\\
Halo HR & 113800 & 70500 & 0.62\\
Disk LR & 10996800 & 10689300 & 0.97\\
Disk LR & 2508300 & 2508300 & 0.82\\
Galaxy Clusters & 72300 & 55300 & 0.76\\
\hspace{0.3cm}\footnotesize{Cluster Galaxies} & 1563400 & 1403700 & 0.90\\
AGN & 763700 & 669800 & 0.88\\
BAO & 15804100 & 12801000 & 0.81\\
\hline
All & 33496600 & 29096300 & 0.87\\
\end{tabular}
\caption{This table summarizes the simulation for 5 years of operations of 4MOST. It is important to note that, for instance, in 10 years, SDSS has collected $2.4$ million spectra, and our estimates are made on very conservative throughput estimates.}
\label{tab:obstarg}
\end{center}
\end{table}
\begin{table}
\begin{center}
\begin{tabular}[c]{lrr}
Channel & Allocated fiber & Unallocated fiber\\
High Resolution & 2969100 & 6430600\\
Low Resolution & 15338000 & 3465500\\
\end{tabular}
\caption{This table summarizes the efficiency of each spectroscopic channel of 4MOST.}
\label{tab:obsefficiency}
\end{center}
\end{table}

\section{Conclusion}
\label{sec:conclusion}
4MOST is a very powerful spectroscopic facility, that will be installed on VISTA in 2019. It will allow the astronomical community to have access to a unique instrument, that will be able to deliver cutting edge data both in the Galactic and the extra-Galactic field. It is currently in a Conceptual Design Phase, and will enter the Preliminary Design Phase early 2015.

Currently, 7 Design Reference Surveys have been identified, that have helped shape the design of the instrument. But there is a lot of room for other Surveys, be they large or small, in order to maximize the scientific output of  the facility. As a survey facility, all data obtained at the telescope will be made public immediately

The 4FS, a very powerful and truly unique tool, allows for the exploration of various observing strategies, to optimize the coverage of the sky, but also allows to test various technical solutions, such as the effect of modifying the number of fibers. As such, it is a virtual 4MOST Instrument, and it is planned to transform it into what will the ETC be.
%\input{referenc}
% BibTeX users please use
%\bibliographystyle{../aa.bst}
%\bibliography{depagne}

% Mac users: please ignore the error message: "! Package natbib Error: Bibliography not compatible with author-year citations."
\end{document}